\documentclass[aps,prl,twocolumn,groupedaddress,showpacs,showkeys,floatfix]{revtex4}

\usepackage{graphicx}
\usepackage{bm}

\def\eq#1{Eq.\ (\ref{#1})}

\def\CK{{\cal K}}

\def\VRGM{V^{\rm RGM}}
\def\VRGMA{V^{\rm RGM}_\alpha}
\def\VRGMB{V^{\rm RGM}_\beta}
\def\VRGMC{V^{\rm RGM}_\gamma}

\begin{document}

\preprint{KUNS****}

\title{
Addendum: Triton and hypertriton binding energies calculated
from $\bm{SU_6}$ quark-model baryon-baryon interactions
}


%
\author{Y. Fujiwara}
\email[]{yfujiwar@scphys.kyoto-u.ac.jp}
\affiliation{
Department of Physics, Kyoto University,
Kyoto 606-8502, Japan
}
\author{Y. Suzuki}
\affiliation{
Department of Physics, and Graduate School
of Science and Technology,
Niigata University, Niigata 950-2181, Japan
}
\author{M. Kohno}
\affiliation{
Physics Division, Kyushu Dental College,
Kitakyushu 803-8580, Japan
}
\author{K. Miyagawa}
\affiliation{
Department of Applied Physics,
Okayama Science University, Okayama 700-0005, Japan
}


\date{\today}

\begin{abstract}
Previously we calculated the binding energies of the
triton and hypertriton, using an $SU_6$ quark-model
interaction derived from a resonating-group method of 
two baryon clusters. In contrast to the previous calculations
employing the energy-dependent interaction kernel, 
we present new results using a renormalized interaction, which 
is now energy independent and reserves all the two-baryon data. 
The new binding energies are slightly smaller than the previous
values. In particular the triton binding energy turns out 
to be 8.14 MeV with a charge-dependence correction of 
the two-nucleon force, 190 keV, being included.
This indicates that about 350 keV is left for the energy
which is to be accounted for by three-body forces. 
\end{abstract}


\maketitle


The QCD-inspired spin-flavor $SU_6$ quark model (QM) for the
baryon-baryon interaction, developed by the Kyoto-Niigata group,
has achieved accurate descriptions
of available $NN$ and $YN$ experimental data \cite{PPNP}.
In particular, the most recent model fss2 gives
in the $NN$ sector accuracy comparable to modern 
realistic meson-exchange potentials.
Since the QM description of the short-range part
is quite different from that of meson-exchange potentials,
it is interesting to apply these interactions
to calculate properties of three-baryon systems,
namely the triton ($\hbox{}^3 \hbox{H}$) and
hypertriton ($\hbox{}^3_\Lambda \hbox{H}$).
For this purpose, we developed in Ref.\,\cite{TRGM,RED} 
a three-cluster equation which employs energy-{\em dependent} 
two-cluster quark-exchange kernels of the resonating-group method (RGM).  
Solving this equation, we obtained the following results for fss2;
the triton binding energy, $B_t=8.519$ MeV \cite{triton,PANIC02},
and the $\Lambda$ separation energy, $B_\Lambda=289$ keV
for $\hbox{}^3_\Lambda \hbox{H}$ \cite{hypt}.
We call this treatment $\varepsilon K$ prescription in the following.

Recently, an important progress is made to apply
the energy-{\em independent} RGM kernel
to the $3\alpha$ system \cite{3alpha} and
other three-cluster systems \cite{rRGM}, through a standard 
procedure \cite{Ti82} of eliminating the energy dependence of the 
RGM kernel. This renormalized kernel 
naturally gives different results in the
application of the QM baryon-baryon interactions to many-body
systems. We will report these new results in this paper. 

In this formulation, the two-cluster RGM kernel $V^{\rm RGM}$ is
expressed as an energy-independent renormalized RGM kernel
\begin{eqnarray}
\VRGM= V_{\rm D}+G+W\ ,
\label{form1}
\end{eqnarray}
where $V_{\rm D}$ is the direct potential, and $G$ is the sum of the
exchange kinetic-energy and interaction kernels.
The kernel $W$ is the term which appears through the elimination 
of the energy-dependence, and it is given by
\begin{eqnarray}
W=\Lambda \frac{1}{\sqrt{1-K}}
h \frac{1}{\sqrt{1-K}}\Lambda-h\ .
\label{form2}
\end{eqnarray}
Here $K$ is the exchange normalization kernel,
$h$ denotes $h_0+V_{\rm D}+G$ with $h_0$ being the relative kinetic-energy
operator, and $\Lambda=1-|u\rangle \langle u|$ is a two-cluster
Pauli projection operator, where $|u \rangle$ is a Pauli forbidden
state satisfying $K|u\rangle=|u\rangle$.
An advantage of this procedure is that the two-cluster RGM equation
takes the form of the usual Schr{\"o}dinger equation in the allowed
model space, and the relative wave function is properly
normalized \cite{Ti82}.
This Schr{\"o}dinger-type equation for the relative wave function 
gives the same asymptotic behavior as the original RGM equation,
thus yielding the same phase shifts and physical observables
for the two-cluster system.
The difference between the previous energy-dependent RGM kernel,
$\VRGM(\varepsilon)=V_{\rm D}+G+\varepsilon K$, and $\VRGM$ in
\eq{form1} is essentially a replacement
of $\Lambda (\varepsilon K) \Lambda$ with $W$.
Here $\varepsilon$ is the two-cluster relative energy measured from its 
threshold, and it was determined in a self-consistent procedure
in the previous $\varepsilon K$ treatment.

In the usual notation, $\alpha$, $\beta$, and $\gamma$,
for three independent pairs of two-cluster subsystems,
the three-cluster equation to be solved reads
\begin{eqnarray}
P \left[\,E-H_0-\VRGMA-\VRGMB-\VRGMC\,\right] P \Psi=0\ ,
\label{form3}
\end{eqnarray}
where $E$ is the three-body energy, $H_0$ is the free three-body
kinetic-energy operator,
and $\VRGMA$ stands for the RGM kernel in \eq{form1} for
the $\alpha$-pair, {\em etc.}
The three-body operator $P$ projects on the Pauli-allowed space
with a proper symmetry of clusters, and it is constructed 
from the orthogonality constraint that each pair of  
two-cluster subsystems is free from any Pauli forbidden states 
\cite{HO74,HO75,Sm74}. This definition
of the three-cluster Pauli-allowed space may not be exactly
equivalent to the standard definition
given by the three-cluster normalization kernel. 
We however employed this orthogonality condition
in the $\varepsilon K$ prescription. See \cite{TRGM,RED} for detail. 
We use the same definition of $P$ in this paper as well.

In the practical applications of the QM baryon-baryon interactions
to the Faddeev formalism,
it is convenient to calculate $W$ in \eq{form2} in the form of
\begin{eqnarray}
W=\CK h+h\CK+\CK h \CK
\label{form4}
\end{eqnarray}
with
\begin{eqnarray}
\CK=\Lambda \left(\frac{1}{\sqrt{1-K}}-1\right)\Lambda\ . 
\label{form5}
\end{eqnarray}
The kernel $\CK$ is calculated in the momentum representation,
by using properties of exchange normalization kernels.
There exists no Pauli forbidden state in the $NN$ interaction
($\Lambda=1$), while one $(0s)$ harmonic-oscillator Pauli forbidden state
appears in the $\Lambda N$--$\Sigma N$ system. 

\begin{table}
\caption{
Three-nucleon bound state properties predicted by fss2 and FSS,
using the energy-independent renormalized RGM kernels.
The column $n$ implies the number of three-nucleon
channels, including the two-nucleon systems up to the total
angular-momentum $J$, $E(\hbox{}^3\hbox{H})$ the ground state
energy, and $\protect\sqrt{\langle r^2\rangle_{\hbox{}^3{\rm H}}}$
($\protect\sqrt{\langle r^2\rangle_{\hbox{}^3{\rm He}}}$) is the charge
rms radius for $\hbox{}^3{\rm H}$ ($\hbox{}^3{\rm He}$) with
the proton and neutron size corrections in \protect\eq{form6}.
The value $\Delta E$ is the energy change
from the $\varepsilon K$ prescription to the present approach.
\label{table1}
}
\renewcommand{\arraystretch}{1.0}
\begin{ruledtabular}
\begin{tabular}{cccccc}
model & $n$
 & $E(\hbox{}^3\hbox{H})$ & $\Delta E$ 
 & $\protect\sqrt{\langle r^2\rangle_{\hbox{}^3{\rm H}}}$
 & $\protect\sqrt{\langle r^2\rangle_{\hbox{}^3{\rm He}}}$ \\
 & & (MeV)  & (keV) & (fm) & (fm) \\
\hline
     &  2 ($S$) & $-7.952$ & $-145$ & 1.80 & 1.95 \\
     &  5 ($S, D$) & $-8.261$ & $-72$ & 1.76 & 1.92 \\
     & 10 ($J \leq 1$) & $-7.962$ &  55 & 1.77 & 1.95 \\
fss2 & 18 ($J \leq 2$) & $-8.228$ & 211 & 1.75 & 1.93 \\
     & 26 ($J \leq 3$) & $-8.313$ &     & 1.75 & 1.92 \\
     & 34 ($J \leq 4$) & $-8.322$ & 192 & 1.75 & 1.92 \\
     & 42 ($J \leq 5$) & $-8.326$ &     & 1.75 & 1.92 \\
     & 50 ($J \leq 6$) & $-8.326$ & 193 & 1.75 & 1.92 \\
\hline
     &  2 ($S$)    & $-7.828$ & $-153$ & 1.82 & 1.97 \\
     &  5 ($S, D$) & $-8.095$ &  $-61$ & 1.78 & 1.95 \\
     & 10 ($J \leq 1$) & $-7.784$ & 125 & 1.80 & 1.98 \\
FSS  & 18 ($J \leq 2$) & $-8.022$ & 320 & 1.78 & 1.96 \\
     & 26 ($J \leq 3$) & $-8.079$ &     & 1.78 & 1.96 \\
     & 34 ($J \leq 4$) & $-8.088$ & 302 & 1.78 & 1.96 \\
     & 42 ($J \leq 5$) & $-8.091$ &     & 1.78 & 1.96 \\
     & 50 ($J \leq 6$) & $-8.091$ & 303 & 1.78 & 1.96 \\
\hline
exp't & & $-8.482$ & & 1.755(86)\footnotemark[1] & 1.959(30)\footnotemark[1] \\
      & &          & & & 1.9642(11)\footnotemark[2] \\
\end{tabular}
\end{ruledtabular}
\footnotetext[1]{Ref.~\onlinecite{Am94}.}
\footnotetext[2]{Ref.~\onlinecite{Mo06}.}
\end{table}

Table \ref{table1} lists three-nucleon bound state properties
predicted by the Faddeev calculations with fss2 and FSS.
The $np$ interaction is employed in the isospin basis.
The momentum discretization points for solving the Faddeev equations
are the same as in Ref.\,\cite{triton}.
The finite size corrections of the nucleons are made through \cite{Ya06}
\begin{eqnarray}
& & \ \hspace{-10mm} \langle r^2\rangle_{\hbox{}^3{\rm H}}
=\left[R_C(\hbox{}^3{\rm H})\right]^2
+(0.8750)^2+2(-0.1161)\ ,\nonumber \\
& & \ \hspace{-10mm} \langle r^2 \rangle_{\hbox{}^3{\rm He}}
=\left[R_C(\hbox{}^3{\rm He})\right]^2
+(0.8750)^2+\frac{1}{2} (-0.1161)\ ,
\label{form6}
\end{eqnarray}
where ${R_C}^2$ stands for the squared charge radius
for the point nucleons.
In order to calculate ${R_C}^2$ from the Faddeev components,
we have improved the previous method using the power series
expansion of the charge form factors.
We have used the second-order numerical differentiation of the
momentum variables in the fifth-order spline interpolation formula,
based on the calculational scheme given in Ref.~\cite{GL82b}.
This approach yields a stable value for the rms radius
within four digits, while in the previous method even the
third digit fluctuates. 
In the present calculation, the Coulomb force and the relativistic
correction terms \cite{Ki88} of the charge current operator
are entirely neglected.

The final fss2 prediction for the triton binding energy is $-8.326$ MeV,
which is 193 keV high, compared with the previous value $-8.519$ MeV.
Since the experimental value is $E^{\rm exp}(\hbox{}^3\hbox{H})
=-8.482$ MeV, the calculated value is higher than
the experiment by 156 keV. In fact, we have to take into account
the effect of the charge dependence of the two-nucleon force,
which is estimated to result in the energy loss
by about 190 keV \cite{Ma89}.
Therefore our calculation concludes that $346$ keV, namely,
about 350 keV is still missing.
In order to compare with the results by the $\varepsilon K$
prescription, we show the energy loss from the previous results
in the column $\Delta E$ in Table \ref{table1}.
In both fss2 and FSS cases, we note that the 5-channel 
energy is already close to the converged value in 
the present approach, whereas the convergence was rather slow in the 
$\varepsilon K$ calculation. 
We will see that this is also the case
in the hypertriton calculation.

We find that the expectation value, $\varepsilon_{NN}
=\langle P\Psi|h_{0\alpha}+V^{\rm RGM}_\alpha|P\Psi \rangle
/\langle P\Psi|P \Psi \rangle$, for the triton
is not very different from the previous value of 
the $\varepsilon K$ prescription.
For example, in the full 50 channel calculations with fss2,
the previous result is $\varepsilon_{NN}=4.492$ MeV,
which is compared with the present result $\varepsilon_{NN}=4.301$ MeV.
The charge rms radii hardly change from the previous values.

\begin{figure}
\includegraphics[width=7.9cm]{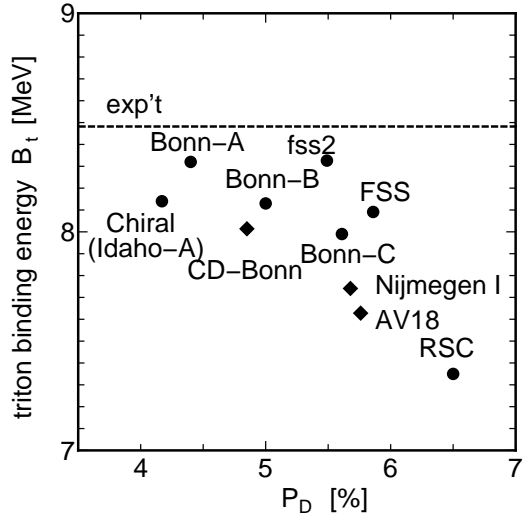}
\caption{\label{tribe}
Calculated $\hbox{}^3\hbox{H}$ binding
energies $B_t$ as a function
of the deuteron $D$-state probability $P_D$.
All the calculations are made in the isospin basis, using the 
$np$ interaction, except for CD-Bonn, Nijmegen I, and AV18
(denoted by black diamonds) which include the effect of
charge-dependence of the interaction.
The calculated values are taken
from \protect\cite{Ma89} (RSC, Bonn-A, B, C),
from \protect\cite{No00} (AV18 \protect\cite{Wi95},
CD-Bonn \protect\cite{Ma01}),
from \protect\cite{No01} (Nijmegen I \protect\cite{St94}),
and from \protect\cite{Em02} (Chiral).
The experimental value, $B_t=8.482$ MeV, is shown by the dashed line.}
\end{figure}

For a realistic calculation of the $\hbox{}^3\hbox{H}$ binding
energy, it is important to use an $NN$ interaction which
reproduces both the proper $D$-state probability $P_D$ of
the deuteron and the effective range parameters
of the $\hbox{}^1S_0$ scattering \cite{Bra88b}.
We show in Fig.~\ref{tribe} the updated plot
of the fss2 and FSS values in the  $B_t=-E(\hbox{}^3\hbox{H})$
{\em vs.} $P_D$ diagram.
We find that fss2 gives a larger binding energy than the modern
realistic meson-exchange potentials like Bonn-C and AV18, 
while the result of FSS is not very far from that of Bonn-C.
It is interesting to note that our QM points are apparently 
off the line on which the data points of the 
modern meson-exchange potentials fall. The five-channel calculation
of the model QCM-A by Takeuchi {\em et al.} \cite{Ta92} gives
almost the same result as Bonn-C.

The results of the hypertriton Faddeev calculations are listed
in Table \ref{table2}.
The $\Lambda$ separation energy of the hypertriton is $B_\Lambda=262$
keV for fss2, which is by 27 keV less than the $\varepsilon K$ value,
289 keV. The corresponding FSS values are 790 keV in the present approach
{\em vs.} 878 keV in the $\varepsilon K$ prescription.
The difference is 88 keV.
So far all the Faddeev calculations, using the energy-independent
renormalized RGM kernels, yield less
binding than the $\varepsilon K$ prescription, as long as the full
model space with enough angular-momenta is taken into account.
Compared with the experimental value,
${B_\Lambda}^{\rm exp}=130 \pm 50~\hbox{keV}$, the fss2 value is
overbound by at least 82 keV. We conclude that the $\Lambda N$ interaction
of fss2 is probably slightly too attractive.
The model FSS has a problem that the attraction of the $\hbox{}^1S_0$
state is too strong, compared with that of the $\hbox{}^3S_1$ state.

From Table \ref{table2}, we again find that
the 15-channel calculation with $S$- and $D$-states only is a good
approximation to the full calculation. We find that
$B_\Lambda=226$ keV for fss2 and 763 keV for FSS in the 15-channel
calculation, and the energy gain to the full calculations
is 36 keV and 27 keV, respectively.
The $NN$ ($\varepsilon_{NN}$) and $\Lambda N$ ($\varepsilon_{\Lambda N}$)
expectation values, and the admixture
of the $\Sigma NN$ component ($P_\Sigma$) are also not much different
from the previous values in the $\varepsilon K$ prescription.
The converged values of $P_\Sigma$ are $0.83 \%$ for fss2 and $1.43
\%$ for FSS, which were previously $0.80 \%$ for fss2 and $1.36 \%$
for FSS, respectively. The decomposition
of the $\varepsilon_{NN}$ value into the kinetic-energy and
potential-energy contributions is $19.034-20.723=-1.689$ MeV
for fss2, which was previously $19.376-21.032=-1.657$ MeV. 
As to the overbinding in the model fss2, we have discussed in
Ref.\,\cite{hypt} that a slight increase of the $\kappa$ meson mass
will improve the fit to the experimental value, without changing
good reproduction of the low-energy $\Lambda N$ cross section data.
If we modify the $\kappa$-meson mass from the original value,
$m_\kappa=936$ MeV, to 995 MeV, we obtain $B_\Lambda=134$ keV
with $P_\Sigma=0.56\,\%$,
which is very close to the NSC89 prediction 
$B_\Lambda=143$ keV with $P_\Sigma=0.5\,\%$ \cite{Mi95,No02}.
The effective range parameters of this modified fss2 interaction
are $a_s=-2.18$ fm, $r_s=3.03$ fm,
and $a_t=-1.78$ fm, $r_t=2.88$ fm. The phase-shift difference is
only $2.2^\circ$ at $p_{\rm lab}=200~\hbox{MeV}/c$.

\begin{table}
\caption{
Results of the $\hbox{}^3_\Lambda \hbox{H}$ Faddeev calculations by
fss2 and FSS, using the energy-independent renormalized RGM kernels.
The momentum discretization points are the same
as in Ref.\,\protect\cite{hypt}.
The calculated deuteron binding energy
is $\varepsilon_d=2.2247$ MeV for fss2 and 2.2561 MeV for FSS
($\varepsilon^{\rm exp}_d=2.2246$ MeV).
The column $n$ implies the number of three-baryon
channels, including the two-baryon systems up to the total
angular-momentum $J$, $E$ the $\hbox{}^3_\Lambda \hbox{H}$
energy measured from the $N+N+\Lambda$ threshold, and $B_\Lambda$
the $\Lambda$ separation energy. The experimental value
is ${B_\Lambda}^{\rm exp}=130 \pm 50~\hbox{keV}$.
The column $P_\Sigma$ shows the squared amplitudes
of the $\Sigma NN$ admixture in percent.
The $\Delta B_\Lambda$ is the energy change from the
previous $\varepsilon K$ prescription to the present
energy-independent renormalized RGM kernels. 
\label{table2}
}
\renewcommand{\arraystretch}{1.0}
\begin{ruledtabular}
\begin{tabular}{cccccc}
model & $n$ & $E$ & $B_\Lambda$
  & $\Delta B_\Lambda$ & $P_\Sigma$ \\
  &   & (MeV) & (keV) & (keV) & ($\%$) \\ 
\hline
     &   6 ($S$)       & $-2.392$ & 167 & $-30$ & 0.566 \\
     &  15 ($SD$)      & $-2.451$ & 226 & $-28$ & 0.775 \\
     &  30 ($J \le 1$) & $-2.404$ & 179 &  $-1$ & 0.679 \\
fss2 &  54 ($J \le 2$) & $-2.467$ & 243 &    31 & 0.792 \\
     &  78 ($J \le 3$) & $-2.483$ & 259 &    27 & 0.824 \\
     & 102 ($J \le 4$) & $-2.486$ & 261 &    27 & 0.828  \\
     & 126 ($J \le 5$) & $-2.487$ & 262 &    27 & 0.830  \\
     & 150 ($J \le 6$) & $-2.487$ & 262 &    27 & 0.830  \\
\hline
     &   6 ($S$)       & $-2.978$ & 722 & $-68$ & 1.251 \\
     &  15 ($SD$   )   & $-3.019$ & 763 & $-53$ & 1.421 \\
     &  30 ($J \le 1$) & $-2.926$ & 670 &    21 & 1.318 \\
FSS  &  54 ($J \le 2$) & $-3.030$ & 774 &    92 & 1.412 \\
     &  78 ($J \le 3$) & $-3.041$ & 785 &    87 & 1.427 \\
     & 102 ($J \le 4$) & $-3.045$ & 789 &    88 & 1.430 \\
     & 126 ($J \le 5$) & $-3.046$ & 790 &    89 & 1.431 \\
     & 150 ($J \le 6$) & $-3.046$ & 790 &    88 & 1.431 \\
\end{tabular}
\end{ruledtabular}
\end{table}

Rather small modification of the present results 
from the previous $\varepsilon K$ prescription is
related to a simple structure of the quark-exchange normalization
kernel $\Lambda K \Lambda$ in the Pauli allowed space.
For the $NN$ interaction, $\Lambda=1$ since there is no Pauli forbidden state.
For the positive-parity states, the largest eigenvalue
of $|K|$ is 1/9 for the $(0s)$ harmonic oscillator state.
Although almost Pauli forbidden states appear
in the $P$-states, such partial waves give
rather minor contributions to the binding energy of the triton.
For the $\Lambda N$--$\Sigma N$ interaction, we have a Pauli forbidden
state classified by the $SU_3$ label $(11)_s$. Once this component is
properly eliminated, the eigenvalues of $\Lambda K \Lambda$ also
become very small. These are the main reasons why the present
treatment by the energy-independent renormalized RGM kernel 
gives the results rather similar to the previous energy-dependent
$\varepsilon K$ prescription. On the contrary, the difference 
between $\Lambda (\varepsilon K) \Lambda$ and $W$ is rather large
in the nuclear cluster systems, which leads to appreciable difference
between these two prescriptions \cite{3alpha,rRGM}.
 
In summary, we have recalculated triton and hypertriton binding
energies in a new semi-microscopic three-cluster equation,
using the energy-independent renormalized RGM kernels
of the quark-model baryon-baryon interactions, fss2 and FSS.
This formulation yields slightly less attractive effect
to the three-baryon systems,
in comparison with the previous energy-dependent treatment
of the two-cluster RGM kernels. 
For the triton result, we conclude that the energy contribution
of the three-nucleon force is not as large as 0.5 - 1 MeV,
predicted by the standard meson-exchange potentials \cite{No00}.
If we compare the fss2 value, 8.326 MeV, with the experimental one,
8.482 MeV, the calculated triton binding energy is too small by 156 keV.
If the charge-dependence correction of 190 keV is further taken
into account, all together 346 keV is still missing.
This lack is, however, almost half of the predictions by the meson-exchange
potentials. For the $\Lambda$ separation energy of the hypertriton,
the overbinding of the model fss2 is slightly reduced.
We still have large ambiguity in $\hbox{}^1S$ and $\hbox{}^3S$
$\Lambda N$ interactions, before further details, such as  
the charge symmetry breaking of $\Lambda p$ and $\Lambda n$
interactions, come into play.
The comparison of the fss2 value, 262 keV, with the experimental one,
$B^{\rm exp}_\Lambda=130 \pm 50$ keV, shows that
the $\hbox{}^1S$ interaction of fss2 is still
too attractive, which can be corrected by choosing a slightly
heavier $\kappa$-meson mass.

\begin{acknowledgments}
This work was supported by Grants-in-Aid for Scientific
Research (C) (Grant Nos.~18540261 and 17540263)
and Bilateral Joint Research Projects (2006-2008)
from the Japan Society for the Promotion
of Science (JSPS).
\end{acknowledgments}

\bibliography{ren}

\begin{thebibliography}{28}
\expandafter\ifx\csname natexlab\endcsname\relax\def\natexlab#1{#1}\fi
\expandafter\ifx\csname bibnamefont\endcsname\relax
  \def\bibnamefont#1{#1}\fi
\expandafter\ifx\csname bibfnamefont\endcsname\relax
  \def\bibfnamefont#1{#1}\fi
\expandafter\ifx\csname citenamefont\endcsname\relax
  \def\citenamefont#1{#1}\fi
\expandafter\ifx\csname url\endcsname\relax
  \def\url#1{\texttt{#1}}\fi
\expandafter\ifx\csname urlprefix\endcsname\relax\def\urlprefix{URL }\fi
\providecommand{\bibinfo}[2]{#2}
\providecommand{\eprint}[2][]{\url{#2}}

\bibitem[{\citenamefont{Fujiwara et~al.}(2007)\citenamefont{Fujiwara, Suzuki,
  and Nakamoto}}]{PPNP}
\bibinfo{author}{\bibfnamefont{Y.}~\bibnamefont{Fujiwara}},
  \bibinfo{author}{\bibfnamefont{Y.}~\bibnamefont{Suzuki}}, \bibnamefont{and}
  \bibinfo{author}{\bibfnamefont{C.}~\bibnamefont{Nakamoto}},
  \bibinfo{journal}{Prog.\ Part.\ Nucl.\ Phys} \textbf{\bibinfo{volume}{58}},
  \bibinfo{pages}{439} (\bibinfo{year}{2007}).

\bibitem[{\citenamefont{Fujiwara
  et~al.}(2002{\natexlab{a}})\citenamefont{Fujiwara, Nemura, Suzuki, Miyagawa,
  and Kohno}}]{TRGM}
\bibinfo{author}{\bibfnamefont{Y.}~\bibnamefont{Fujiwara}},
  \bibinfo{author}{\bibfnamefont{H.}~\bibnamefont{Nemura}},
  \bibinfo{author}{\bibfnamefont{Y.}~\bibnamefont{Suzuki}},
  \bibinfo{author}{\bibfnamefont{K.}~\bibnamefont{Miyagawa}}, \bibnamefont{and}
  \bibinfo{author}{\bibfnamefont{M.}~\bibnamefont{Kohno}},
  \bibinfo{journal}{Prog.\ Theor.\ Phys.} \textbf{\bibinfo{volume}{107}},
  \bibinfo{pages}{745} (\bibinfo{year}{2002}{\natexlab{a}}).

\bibitem[{\citenamefont{Fujiwara
  et~al.}(2002{\natexlab{b}})\citenamefont{Fujiwara, Suzuki, Miyagawa, Kohno,
  and Nemura}}]{RED}
\bibinfo{author}{\bibfnamefont{Y.}~\bibnamefont{Fujiwara}},
  \bibinfo{author}{\bibfnamefont{Y.}~\bibnamefont{Suzuki}},
  \bibinfo{author}{\bibfnamefont{K.}~\bibnamefont{Miyagawa}},
  \bibinfo{author}{\bibfnamefont{M.}~\bibnamefont{Kohno}}, \bibnamefont{and}
  \bibinfo{author}{\bibfnamefont{H.}~\bibnamefont{Nemura}},
  \bibinfo{journal}{Prog.\ Theor.\ Phys.} \textbf{\bibinfo{volume}{107}},
  \bibinfo{pages}{993} (\bibinfo{year}{2002}{\natexlab{b}}).

\bibitem[{\citenamefont{Fujiwara
  et~al.}(2002{\natexlab{c}})\citenamefont{Fujiwara, Miyagawa, Kohno, Suzuki,
  and Nemura}}]{triton}
\bibinfo{author}{\bibfnamefont{Y.}~\bibnamefont{Fujiwara}},
  \bibinfo{author}{\bibfnamefont{K.}~\bibnamefont{Miyagawa}},
  \bibinfo{author}{\bibfnamefont{M.}~\bibnamefont{Kohno}},
  \bibinfo{author}{\bibfnamefont{Y.}~\bibnamefont{Suzuki}}, \bibnamefont{and}
  \bibinfo{author}{\bibfnamefont{H.}~\bibnamefont{Nemura}},
  \bibinfo{journal}{Phys.\ Rev.\ C} \textbf{\bibinfo{volume}{66}},
  \bibinfo{pages}{021001} (\bibinfo{year}{2002}{\natexlab{c}}).

\bibitem[{\citenamefont{Fujiwara et~al.}(2003)\citenamefont{Fujiwara, Miyagawa,
  Suzuki, Kohno, and Nemura}}]{PANIC02}
\bibinfo{author}{\bibfnamefont{Y.}~\bibnamefont{Fujiwara}},
  \bibinfo{author}{\bibfnamefont{K.}~\bibnamefont{Miyagawa}},
  \bibinfo{author}{\bibfnamefont{Y.}~\bibnamefont{Suzuki}},
  \bibinfo{author}{\bibfnamefont{M.}~\bibnamefont{Kohno}}, \bibnamefont{and}
  \bibinfo{author}{\bibfnamefont{H.}~\bibnamefont{Nemura}},
  \bibinfo{journal}{Nucl.\ Phys.} \textbf{\bibinfo{volume}{A721}},
  \bibinfo{pages}{983c} (\bibinfo{year}{2003}).

\bibitem[{\citenamefont{Fujiwara et~al.}(2004)\citenamefont{Fujiwara, Miyagawa,
  Kohno, and Suzuki}}]{hypt}
\bibinfo{author}{\bibfnamefont{Y.}~\bibnamefont{Fujiwara}},
  \bibinfo{author}{\bibfnamefont{K.}~\bibnamefont{Miyagawa}},
  \bibinfo{author}{\bibfnamefont{M.}~\bibnamefont{Kohno}}, \bibnamefont{and}
  \bibinfo{author}{\bibfnamefont{Y.}~\bibnamefont{Suzuki}},
  \bibinfo{journal}{Phys.\ Rev.\ C} \textbf{\bibinfo{volume}{70}},
  \bibinfo{pages}{024001} (\bibinfo{year}{2004}).

\bibitem[{\citenamefont{Suzuki et~al.}()\citenamefont{Suzuki, Matsumura, Orabi,
  Fujiwara, Descouvemont, Theeten, and Baye}}]{3alpha}
\bibinfo{author}{\bibfnamefont{Y.}~\bibnamefont{Suzuki}},
  \bibinfo{author}{\bibfnamefont{H.}~\bibnamefont{Matsumura}},
  \bibinfo{author}{\bibfnamefont{M.}~\bibnamefont{Orabi}},
  \bibinfo{author}{\bibfnamefont{Y.}~\bibnamefont{Fujiwara}},
  \bibinfo{author}{\bibfnamefont{P.}~\bibnamefont{Descouvemont}},
  \bibinfo{author}{\bibfnamefont{M.}~\bibnamefont{Theeten}}, \bibnamefont{and}
  \bibinfo{author}{\bibfnamefont{D.}~\bibnamefont{Baye}}, \eprint{to be
  published in Phys. Lett. B (2007)}.

\bibitem[{\citenamefont{Theeten et~al.}()\citenamefont{Theeten, Matsumura,
  Orabi, Descouvemont, Fujiwara, and Suzuki}}]{rRGM}
\bibinfo{author}{\bibfnamefont{M.}~\bibnamefont{Theeten}},
  \bibinfo{author}{\bibfnamefont{H.}~\bibnamefont{Matsumura}},
  \bibinfo{author}{\bibfnamefont{M.}~\bibnamefont{Orabi}},
  \bibinfo{author}{\bibfnamefont{D.~B.~P.} \bibnamefont{Descouvemont}},
  \bibinfo{author}{\bibfnamefont{Y.}~\bibnamefont{Fujiwara}}, \bibnamefont{and}
  \bibinfo{author}{\bibfnamefont{Y.}~\bibnamefont{Suzuki}}, \eprint{submitted
  to Phys. Rev. C (2007)}.

\bibitem[{\citenamefont{Timm et~al.}(1982)\citenamefont{Timm, Fiebig, and
  Friedrich}}]{Ti82}
\bibinfo{author}{\bibfnamefont{W.}~\bibnamefont{Timm}},
  \bibinfo{author}{\bibfnamefont{H.}~\bibnamefont{Fiebig}}, \bibnamefont{and}
  \bibinfo{author}{\bibfnamefont{H.}~\bibnamefont{Friedrich}},
  \bibinfo{journal}{Phys.\ Rev.\ C} \textbf{\bibinfo{volume}{25}},
  \bibinfo{pages}{79} (\bibinfo{year}{1982}).

\bibitem[{\citenamefont{Horiuchi}(1974)}]{HO74}
\bibinfo{author}{\bibfnamefont{H.}~\bibnamefont{Horiuchi}},
  \bibinfo{journal}{Prog.\ Theor.\ Phys.} \textbf{\bibinfo{volume}{51}},
  \bibinfo{pages}{1266} (\bibinfo{year}{1974}).

\bibitem[{\citenamefont{Horiuchi}(1975)}]{HO75}
\bibinfo{author}{\bibfnamefont{H.}~\bibnamefont{Horiuchi}},
  \bibinfo{journal}{Prog.\ Theor.\ Phys.} \textbf{\bibinfo{volume}{53}},
  \bibinfo{pages}{447} (\bibinfo{year}{1975}).

\bibitem[{\citenamefont{Smirnov et~al.}(1974)\citenamefont{Smirnov, Obukhovsky,
  Tchuvil'sky, and Neudatchin}}]{Sm74}
\bibinfo{author}{\bibfnamefont{Y.~F.} \bibnamefont{Smirnov}},
  \bibinfo{author}{\bibfnamefont{I.~T.} \bibnamefont{Obukhovsky}},
  \bibinfo{author}{\bibfnamefont{Y.~M.} \bibnamefont{Tchuvil'sky}},
  \bibnamefont{and} \bibinfo{author}{\bibfnamefont{V.~G.}
  \bibnamefont{Neudatchin}}, \bibinfo{journal}{Nucl.\ Phys.}
  \textbf{\bibinfo{volume}{A235}}, \bibinfo{pages}{289} (\bibinfo{year}{1974}).

\bibitem[{\citenamefont{{Amroun, {\em et al.}}}(1994)}]{Am94}
\bibinfo{author}{\bibfnamefont{A.}~\bibnamefont{{Amroun, {\em et al.}}}},
  \bibinfo{journal}{Nucl.\ Phys.} \textbf{\bibinfo{volume}{A579}},
  \bibinfo{pages}{596} (\bibinfo{year}{1994}).

\bibitem[{\citenamefont{Morton et~al.}(2006)\citenamefont{Morton, Wu, and
  Drake}}]{Mo06}
\bibinfo{author}{\bibfnamefont{D.~C.} \bibnamefont{Morton}},
  \bibinfo{author}{\bibfnamefont{Q.}~\bibnamefont{Wu}}, \bibnamefont{and}
  \bibinfo{author}{\bibfnamefont{G.~W.~F.} \bibnamefont{Drake}},
  \bibinfo{journal}{Phys.\ Rev.\ A} \textbf{\bibinfo{volume}{73}},
  \bibinfo{pages}{034502} (\bibinfo{year}{2006}).

\bibitem[{\citenamefont{{Yao, {\em et al.}}}(2006)}]{Ya06}
\bibinfo{author}{\bibfnamefont{W.~M.} \bibnamefont{{Yao, {\em et al.}}}},
  \bibinfo{journal}{J.\ Phys.\ G} \textbf{\bibinfo{volume}{33}},
  \bibinfo{pages}{1} (\bibinfo{year}{2006}).

\bibitem[{\citenamefont{Gl{\" o}ckle}(1982)}]{GL82b}
\bibinfo{author}{\bibfnamefont{W.}~\bibnamefont{Gl{\" o}ckle}},
  \bibinfo{journal}{Nucl.\ Phys.} \textbf{\bibinfo{volume}{A381}},
  \bibinfo{pages}{343} (\bibinfo{year}{1982}).

\bibitem[{\citenamefont{Kim et~al.}(1988)\citenamefont{Kim, Kim, Klepacki,
  Brandenburg, Harper, and Machleidt}}]{Ki88}
\bibinfo{author}{\bibfnamefont{K.~T.} \bibnamefont{Kim}},
  \bibinfo{author}{\bibfnamefont{Y.~E.} \bibnamefont{Kim}},
  \bibinfo{author}{\bibfnamefont{D.~J.} \bibnamefont{Klepacki}},
  \bibinfo{author}{\bibfnamefont{R.~A.} \bibnamefont{Brandenburg}},
  \bibinfo{author}{\bibfnamefont{E.~P.} \bibnamefont{Harper}},
  \bibnamefont{and}
  \bibinfo{author}{\bibfnamefont{R.}~\bibnamefont{Machleidt}},
  \bibinfo{journal}{Phys.\ Rev.\ C} \textbf{\bibinfo{volume}{38}},
  \bibinfo{pages}{2366} (\bibinfo{year}{1988}).

\bibitem[{\citenamefont{Machleidt}(1989)}]{Ma89}
\bibinfo{author}{\bibfnamefont{R.}~\bibnamefont{Machleidt}},
  \bibinfo{journal}{{\em Adv. Nucl. Phys.}} \textbf{\bibinfo{volume}{19}},
  \bibinfo{pages}{189} (\bibinfo{year}{1989}).

\bibitem[{\citenamefont{Nogga et~al.}(2000{\natexlab{a}})\citenamefont{Nogga,
  Kamada, and Gl{\" o}ckle}}]{No00}
\bibinfo{author}{\bibfnamefont{A.}~\bibnamefont{Nogga}},
  \bibinfo{author}{\bibfnamefont{H.}~\bibnamefont{Kamada}}, \bibnamefont{and}
  \bibinfo{author}{\bibfnamefont{W.}~\bibnamefont{Gl{\" o}ckle}},
  \bibinfo{journal}{Phys.\ Rev.\ Lett.} \textbf{\bibinfo{volume}{85}},
  \bibinfo{pages}{944} (\bibinfo{year}{2000}{\natexlab{a}}).

\bibitem[{\citenamefont{Wiringa et~al.}(1995)\citenamefont{Wiringa, Stoks, and
  Schiavilla}}]{Wi95}
\bibinfo{author}{\bibfnamefont{R.~B.} \bibnamefont{Wiringa}},
  \bibinfo{author}{\bibfnamefont{V.~G.~J.} \bibnamefont{Stoks}},
  \bibnamefont{and}
  \bibinfo{author}{\bibfnamefont{R.}~\bibnamefont{Schiavilla}},
  \bibinfo{journal}{Phys.\ Rev.\ C} \textbf{\bibinfo{volume}{51}},
  \bibinfo{pages}{38} (\bibinfo{year}{1995}).

\bibitem[{\citenamefont{Machleidt}(2001)}]{Ma01}
\bibinfo{author}{\bibfnamefont{R.}~\bibnamefont{Machleidt}},
  \bibinfo{journal}{Phys.\ Rev.\ C} \textbf{\bibinfo{volume}{63}},
  \bibinfo{pages}{024001} (\bibinfo{year}{2001}).

\bibitem[{\citenamefont{Nogga}()}]{No01}
\bibinfo{author}{\bibfnamefont{A.}~\bibnamefont{Nogga}}, \eprint{Ph. D. thesis,
  Ruhr-Universit{\"a}t Bochum (2001)}.

\bibitem[{\citenamefont{Stoks et~al.}(1994)\citenamefont{Stoks, Klomp,
  Terheggen, and de~Swart}}]{St94}
\bibinfo{author}{\bibfnamefont{V.~G.~J.} \bibnamefont{Stoks}},
  \bibinfo{author}{\bibfnamefont{R.~A.~M.} \bibnamefont{Klomp}},
  \bibinfo{author}{\bibfnamefont{C.~P.~F.} \bibnamefont{Terheggen}},
  \bibnamefont{and} \bibinfo{author}{\bibfnamefont{J.~J.}
  \bibnamefont{de~Swart}}, \bibinfo{journal}{Phys.\ Rev.\ C}
  \textbf{\bibinfo{volume}{49}}, \bibinfo{pages}{2950} (\bibinfo{year}{1994}).

\bibitem[{\citenamefont{Entem and Machleit}(2002)}]{Em02}
\bibinfo{author}{\bibfnamefont{D.~R.} \bibnamefont{Entem}} \bibnamefont{and}
  \bibinfo{author}{\bibfnamefont{R.}~\bibnamefont{Machleit}},
  \bibinfo{journal}{Phys.\ Lett.} \textbf{\bibinfo{volume}{B524}},
  \bibinfo{pages}{93} (\bibinfo{year}{2002}).

\bibitem[{\citenamefont{Brandenburg et~al.}(1988)\citenamefont{Brandenburg,
  Chulick, Machleidt, Picklesimer, and Thaler}}]{Bra88b}
\bibinfo{author}{\bibfnamefont{R.~A.} \bibnamefont{Brandenburg}},
  \bibinfo{author}{\bibfnamefont{G.~S.} \bibnamefont{Chulick}},
  \bibinfo{author}{\bibfnamefont{R.}~\bibnamefont{Machleidt}},
  \bibinfo{author}{\bibfnamefont{A.}~\bibnamefont{Picklesimer}},
  \bibnamefont{and} \bibinfo{author}{\bibfnamefont{R.~M.}
  \bibnamefont{Thaler}}, \bibinfo{journal}{Phys.\ Rev.\ C}
  \textbf{\bibinfo{volume}{37}}, \bibinfo{pages}{1245} (\bibinfo{year}{1988}).

\bibitem[{\citenamefont{Takeuchi et~al.}(1992)\citenamefont{Takeuchi, Cheon,
  and Redish}}]{Ta92}
\bibinfo{author}{\bibfnamefont{S.}~\bibnamefont{Takeuchi}},
  \bibinfo{author}{\bibfnamefont{T.}~\bibnamefont{Cheon}}, \bibnamefont{and}
  \bibinfo{author}{\bibfnamefont{E.~F.} \bibnamefont{Redish}},
  \bibinfo{journal}{Phys.\ Rev.\ Lett.} \textbf{\bibinfo{volume}{B280}},
  \bibinfo{pages}{175} (\bibinfo{year}{1992}).

\bibitem[{\citenamefont{Miyagawa et~al.}(1995)\citenamefont{Miyagawa, Kamada,
  Gl{\"o}ckle, and Stoks}}]{Mi95}
\bibinfo{author}{\bibfnamefont{K.}~\bibnamefont{Miyagawa}},
  \bibinfo{author}{\bibfnamefont{H.}~\bibnamefont{Kamada}},
  \bibinfo{author}{\bibfnamefont{W.}~\bibnamefont{Gl{\"o}ckle}},
  \bibnamefont{and} \bibinfo{author}{\bibfnamefont{V.}~\bibnamefont{Stoks}},
  \bibinfo{journal}{Phys.\ Rev.\ C} \textbf{\bibinfo{volume}{51}},
  \bibinfo{pages}{2905} (\bibinfo{year}{1995}).

\bibitem[{\citenamefont{Nogga et~al.}(2000{\natexlab{b}})\citenamefont{Nogga,
  Kamada, and Gl{\" o}ckle}}]{No02}
\bibinfo{author}{\bibfnamefont{A.}~\bibnamefont{Nogga}},
  \bibinfo{author}{\bibfnamefont{H.}~\bibnamefont{Kamada}}, \bibnamefont{and}
  \bibinfo{author}{\bibfnamefont{W.}~\bibnamefont{Gl{\" o}ckle}},
  \bibinfo{journal}{Phys.\ Rev.\ Lett.} \textbf{\bibinfo{volume}{88}},
  \bibinfo{pages}{172501} (\bibinfo{year}{2000}{\natexlab{b}}).

\end{thebibliography}

\end{document}